\documentclass[conference]{IEEEtran}

\usepackage{balance}

\ifCLASSOPTIONcompsoc
  \usepackage[nocompress]{cite}
\else
  \usepackage{cite}
\fi
\ifCLASSINFOpdf
\else
\fi

\usepackage[
  pdftitle={Can Pairwise Testing Perform Comparably to Manually Handcrafted Testing Carried Out by Industrial Engineers?}, 
  pdfauthor={Peter Charbachi, Linus Eklund \& Eduard Enoiu}, 
  colorlinks=true, 
  allcolors=black, 
  raiselinks=false,
  bookmarksopen=true]{hyperref}
\usepackage[table,usenames,dvipsnames]{xcolor}
\usepackage{paralist}
\usepackage{graphicx}
\usepackage[utf8]{inputenc}
\usepackage{multirow}
\usepackage{hhline}

\newcommand{\iec}{{\sc IEC 61131-3}}
\newcommand{\ctt}{{\sc Seafox}} 

\usepackage[framemethod=tikz]{mdframed}
\mdfdefinestyle{style1}{leftmargin=1cm,rightmargin=0.5cm,skipbelow=0.5cm,skipabove=0.3cm,innertopmargin=4pt}

\begin{document}
%
\title{Can Pairwise Testing Perform Comparably to Manually Handcrafted Testing Carried Out by Industrial Engineers?}
%
%
%
%
\author{
    \IEEEauthorblockN{Peter~Charbachi\IEEEauthorrefmark{1}, Linus~Eklund\IEEEauthorrefmark{1}, and Eduard~Enoiu\IEEEauthorrefmark{1}}
    \IEEEauthorblockA{\IEEEauthorrefmark{1}M\"alardalen University, V\"aster{\aa}s,  Sweden.}
}
\markboth{Journal of \LaTeX\ Class Files,~Vol.~14, No.~8, August~2015}%
{Shell \MakeLowercase{\textit{et al.}}: Bare Advanced Demo of IEEEtran.cls for IEEE Computer Society Journals}
%



\IEEEtitleabstractindextext{%
\begin{abstract}
Testing is an important activity in engineering of industrial software. For such software, testing is usually performed manually by handcrafting test suites based on specific design techniques and domain-specific experience. To support developers in testing, different approaches for producing good test suites have been proposed. In the last couple of years combinatorial testing has been explored with the goal of automatically combining the input values of the software based on a certain strategy. Pairwise testing is a combinatorial technique used to generate test suites by varying the values of each pair of input parameters to a system until all possible combinations of those parameters are created. There is some evidence suggesting that these kinds of techniques are efficient and relatively good at detecting software faults. Unfortunately, there is little experimental evidence on the comparison of these combinatorial testing techniques with, what is perceived as, rigorous manually handcrafted testing. In this study we compare pairwise test suites with test suites created manually by engineers for 45 industrial programs. The test suites were evaluated in terms of fault detection, code coverage and number of tests. The results of this study show that pairwise testing, while useful for achieving high code coverage and fault detection for the majority of the programs, is almost as effective in terms of fault detection as manual testing. The results also suggest that pairwise testing is just as good as manual testing at fault detection for 64\% of the programs. 
\end{abstract}

}

\maketitle

\IEEEdisplaynontitleabstractindextext

%
\IEEEpeerreviewmaketitle

\section{Introduction}
\label{sec:introduction}

%
%
%
%
\IEEEPARstart{S}{oftware} testing \cite{ammann2008introduction} is an important activity used for verification and validation by observing the software, executed using a set of test inputs. In practice, engineers are creating these inputs based on different test goals and test design techniques (e.g., specification-based, random, combinatorial, code coverage-based). These techniques have so far been performed manually or semi-automatically with respect to distinct software development activities (i.e, unit and integration testing). With the emerging use of large complex software products, the traditional way of testing software has changed; engineers need to deliver high-quality software while devoting less time for properly testing the software. 

In practice, test suites are still created manually by handcrafting them using specific test design techniques and domain-specific experience. Although the automatic or semi-automatic creation of test suites has been the focus of a great deal of research, manual testing is still widely used \cite{andersson2002verification,beer2008role} in the software development industry. However, over the past few decades, several test design techniques \cite{ammann2008introduction} have been proposed for the creation of test suites with less effort. Combinatorial testing \cite{cohen1996combinatorial} is a technique that creates test inputs based on combinations among the input values. Pairwise testing is an approach to combinatorial testing that generates a test suite which covers each combination of value pairs at least once. There is some evidence\cite{borazjany2012combinatorial, hagar2015introducing} suggesting that pairwise testing is efficient and effective at detecting software faults. However, even if pairwise techniques have been found useful and applicable in industrial applications, the experimental evidence regarding its effectiveness in practice is still limited. 

In this paper we compare pairwise testing and manual testing performed by industrial engineers on industrial software created using the {\iec} programming language \cite{international3iec} that runs on Programmable Logic Controllers (PLCs). The paper makes the following contributions:
\begin{itemize}
\item Empirical evidence showing that pairwise testing achieves marginally lower levels of code coverage while in the same time using more tests cases on average than manual testing performed by industrial engineers. 
\item Results showing that manual testing is not significantly better at finding faults than pairwise testing. Our paper suggests that pairwise testing is just as good in fault detection as manual testing for 64\% of the programs considered.
\item A discussion of the implications of these results for test engineers and researchers.
\end{itemize}

\section{Background}
\label{sec:background}

This paper describes a case study evaluating pairwise testing when used on PLC industrial programs implemented in the IEC 61131-3 FBD language. In this section, we provide a background on PLC industrial software and pairwise testing. According to Ammann and Offutt \cite{ammann2008introduction}, a test case is a set of inputs, expected outputs and actual outputs executed on the specified program. A test suite is a set of ordered test cases. Throughout the paper, we will use the terms test case and test suite in this way.


\subsection{Programmable Logic Controllers}
\label{subsec:plc}

A Programmable Logic Controller (PLC) is a computer system containing a processor, a memory, and a communication bus. PLCs \cite{lewis1998programming} have a programmable memory for storing the software used for expressing logical behaviour, timing and input/output control, networking and data processing. Safety-critical industrial systems implemented using PLCs are used in many applications \cite{parnas1990evaluation} such as transportation, robotics, nuclear and pharmaceutical. Software running on a PLC execute in a loop called scan cycle, in which the iteration follows the “read-execute-write” semantics. The PLC reads the input signals, computes the logical behaviour without interruption and updates its output signals \cite{donandt1989improving}. 

\begin{figure}
\includegraphics[width=\columnwidth]{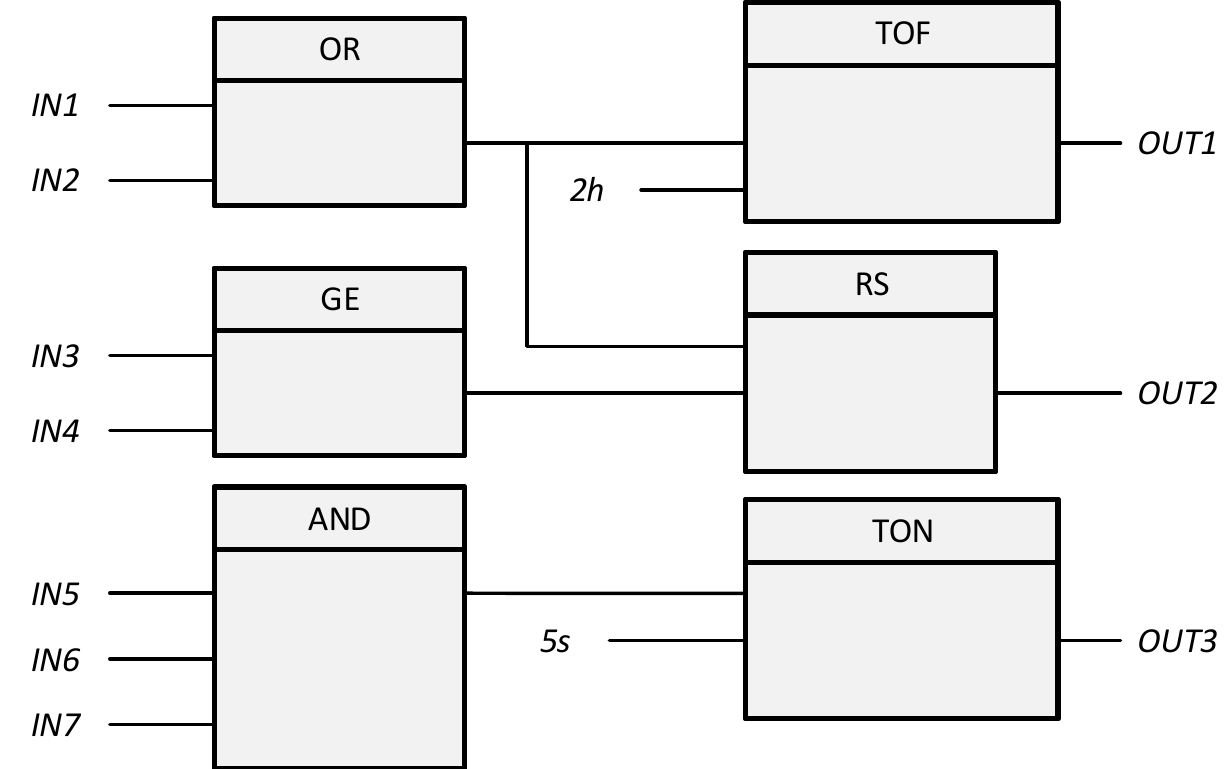}
\caption{A PLC program with seven inputs and three outputs written
using the FBD IEC 61131-3 programming language.}
\label{figure:fbd}
\end{figure}

Programming a PLC differs from general-purpose computers; the PLC software follows a standardized programming paradigm: the IEC (International Electronical Commission) 61131-3 standard \cite{international3iec}. IEC 61131-3 is a popular programming language standard for PLCs used in industrial practice. As shown in Figure \ref{figure:fbd}, computational blocks in an IEC 61131-3 program can be represented in a Function Block Diagram (FBD). This program contains
predefined logical and/or stateful blocks (i.e., OR, RS, TOF, GE, AND and TON in Figure \ref{figure:fbd}) and signals (i.e., connections) between blocks representing the whole behavior of an FBD program. PLC software contains a particular type of blocks named timers that are used to activate or deactivate an output signal after a specific time interval ~\cite{lewis1998programming}. A timer block (e.g., TON and TOF in Figure \ref{figure:fbd}) keeps track of the number of times its input is either true or false and outputs different signals. The IEC 61131-3 standard contains four other programming languages: Instruction List (IL), Structured Text (ST), Ladder Diagram (LD) and Sequential Function Chart (SFC)~\cite{bolton2015programmable}. For more details on PLC programming and FBDs we refer the reader to the work of John et al. \cite{john2010iec}.

\subsection{Pairwise Testing}
\label{subsec:auto}
The process of test generation is that of finding suitable test inputs
using a certain test goal that guides the search in an algorithmic way \cite{ammann2008introduction}. Many algorithms and techniques \cite{orso2014software} for test generation have been proposed. One such technique is combinatorial testing which is used to reveal faults
caused by interactions between input parameters inside a software program. Such techniques design test cases by combining different input parameters based on a combinatorial strategy. Grindal et al. \cite{grindal2006evaluation} surveyed several strategies used for combinatorial testing (e.g., each-used, pair-wise, t-wise, base choice). One of the most commonly used strategy is pairwise (also known as two-way) testing in which each combination for all possible pairs of input parameters are covered by at least one test case. Several empirical studies \cite{borazjany2012combinatorial, hagar2015introducing,li2016applying,grindal2006evaluation} on the use of combinatorial testing for industrial software have been reported and showed that pairwise testing is a very effective technique. In this paper we seek to investigate the use of pairwise testing for industrial control software and compare this technique with manual testing performed by industrial engineers.



\section{Related Work}
Most studies concerning pairwise testing and related to the work included in this paper have focused on how to generate tests as quickly as possible, measure the code coverage score and/or compare with other combinatorial criteria \cite{grindal2006evaluation} or with random tests \cite{ghandehari2014empirical,schroeder2004comparing}. For example, Cohen et al.~\cite{cohen1996combinatorial} found that pairwise generated test suites can achieve 90\% block code coverage. These test suites where generated by the AETG tool. The same tool was used by Burr and Young~\cite{burr1998combinatorial} in a different study. In this paper, pairwise testing achieved 93\% block coverage on average. In addition, Vilkomir and Anderson~\cite{vilkomir2015relationship} showed that pairwise test suites could achieve 77\% MC/DC code coverage.  

Other studies \cite{cohen1994automatic,dalal1998model,sampath2012improving} have reported the use of pairwise testing on real systems and how it can help in the detection of additional bugs when compared to standard test techniques. On the other hand, a few other studies compared manual with pairwise testing \cite{ellims2008effectiveness,ellims2007aetg} and the results suggest that pairwise testing is not able to detect more faults than manually created tests. These results encouraged our interest in investigating on a larger case study how manual testing performed by industrial engineers compares to pairwise testing for industrial software systems. Is there any compelling evidence on how pairwise test suites compare with rigorously handcrafted test suites in terms of test effectiveness? 

\begin{figure*}[!ht]
\vspace{-30mm}
\centering
\includegraphics[width=0.65\linewidth]{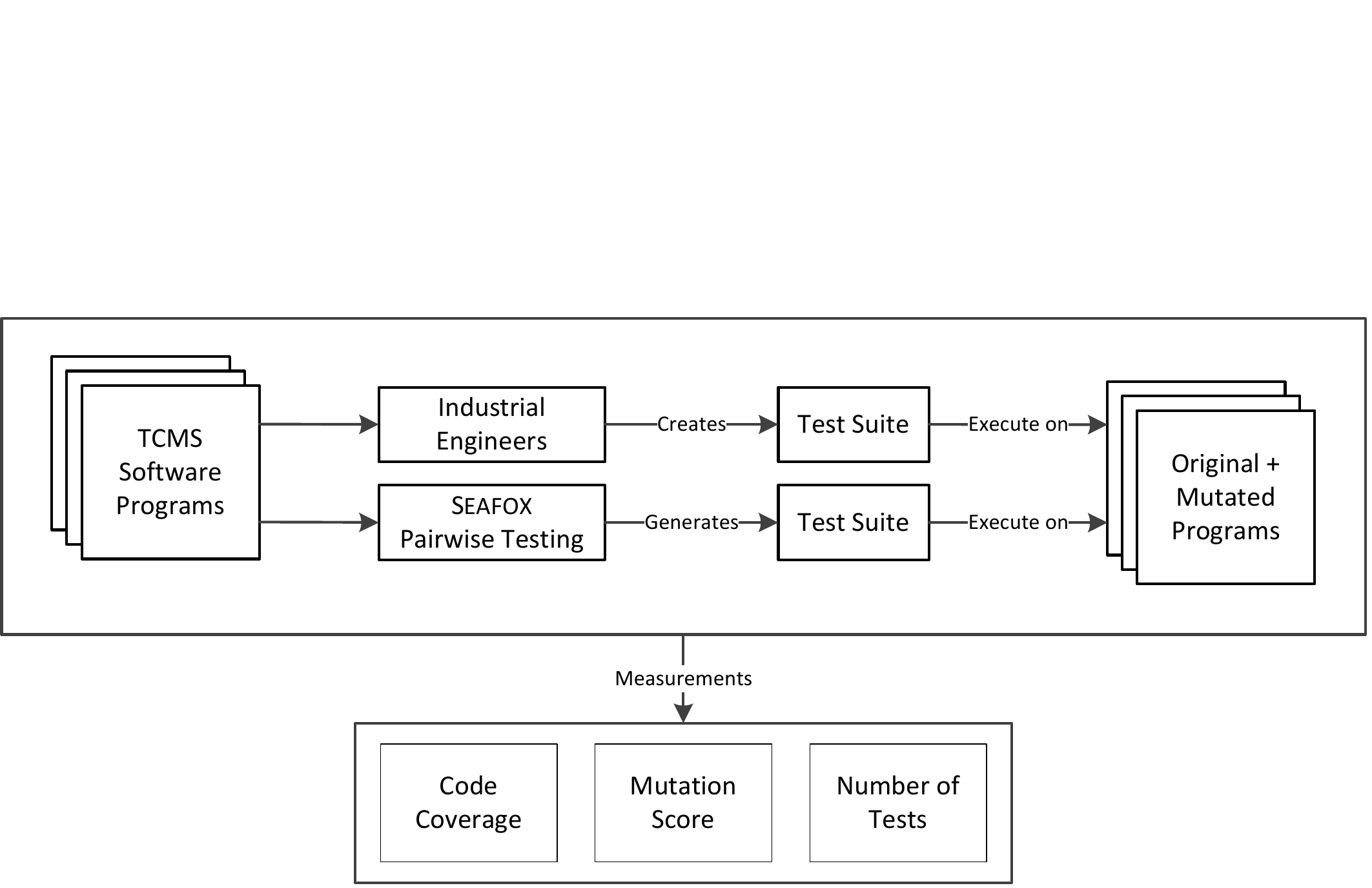}
\caption[Methodology overview]{Overview of the experimental method used to perform the case study. For each program in the Train Control Management System (TCMS), test suites are created manually by industrial engineers, generated by {\ctt} for pairwise testing, and executed on both the original and the mutated programs in order to collect scores for code coverage, mutation and number of tests.}
\label{figure:methodology}
\end{figure*}

\section{Method}
\label{sec:method}
The goal of this paper is to study the comparison between manual test suites created by industrial engineers and automatically generated test suites using a generation tool for pairwise testing in terms of efficiency and effectiveness of testing. To achieve this goal, we designed a case study (mirrored in Figure \ref{figure:methodology}) using industrial software programs from an already developed train control management system to answer the following research questions:
%
%
\begin{itemize}
\item {\it RQ1: Are pairwise generated test suites able to cover more code than test suites manually created by industrial engineers?}
\item {\it RQ2: Are pairwise generated test suites able to detect more faults than test suites manually created by industrial engineers?}
\item {\it RQ3: Is the size of pairwise generated test suites smaller than those manually created by industrial engineers?}
\end{itemize}
  
For each selected program, we executed the test suites produced by both manual testing and pairwise testing and collected the following measures: branch coverage in terms of achieved code coverage, the number of generated test cases and the mutation score as a proxy for fault detection. In order to calculate the mutation score, each test suite was executed on the mutated versions of the original program to determine whether it detects the injected fault. 
This section describes the design of our case study, including the subject programs, the evaluation metrics and the test generation and selection.


\subsection{Subject Programs}
\label{subsec:sut}
Our case study uses an industrial safety-critical system developed by Bombardier Transportation Sweden AB, a large-scale company developing and manufacturing railway equipment. The system is a train control management software ({\sc TCMS}) that has been in development for several years and is tested according to safety standards and regulations. TCMS is a control system containing both software and hardware components in charge of the safety-related functionality of the train and is used by Bombardier Transportation Sweden AB for the control and communication functions in high speed trains. These functions are developed as software programs for PLCs using the Function Block Diagram (FBD) {\iec} graphical programming language \cite{international3iec}. Programs in TCMS are developed in a graphical development environment, compiled into PLC code and saved in standardized PLCOpen XML \footnote{http://www.plcopen.org/} containing structural and behavioural declarations. 

We selected the subject programs for our case study by investigating the TCMS programs provided by Bombardier Transportation Sweden AB. We identified 53 programs and excluded eight programs due to the following reasons: one program contained only one input parameter, for another program the test generation got a memory exception, while for the six remaining programs the test execution failed due to wrong parameter ranges that resulted in an execution exception. Our final set of subjects contains 45 programs. These programs contain nine input parameters and 1076 Lines of XML Source Code (LOC) on average per program. The studied programs were already thoroughly manually tested and are currently used in an operational train. 



\subsection{Test Case Creation}
\label{subsec:testcase}

We used manual test suites created by industrial engineers working at Bombardier Transportation Sweden AB. These manual test suites were obtained by using a post-mortem analysis of the data provided. In testing these programs, engineers perform testing according to specific safety standards and regulations. Specification-based testing is used by engineers to manually create test suites as this is mandated by the EN50128 standard \cite{en200150128}. The test suites collected in this study were based on functional requirement specifications written in a natural language. 

In addition, we generate pairwise test suites using {\ctt} \cite{peter_2017_439253}. {\ctt} is the only available combinatorial test suite generation tool for {\iec} control software. {\ctt} is open source
software and is available at https://github.com/CharByte/SEAFOX \footnote{For more details on the {\ctt} tool we refer the reader to the work of Charbachi and Eklund \cite{thesischa}.}.

{\ctt} supports the generation of test suites using pairwise, base choice and random strategies. For pairwise generation, {\ctt} uses the IPOG algorithm as well as a first pick tie breaker\cite{lei2008ipog}. {\ctt} was used in this study as it supports as input a standard PLCOpen XML implementation of the programs. 
A developer using {\ctt} can automatically generate test suites needed for a given {\iec} program after manually providing the input parameter range information based on the defined behaviour written in the specification.


In order to collect realistic data, we asked one test engineer from Bombardier Transportation, responsible for testing {\iec} software used in this study, to identify the range values for each input parameters and constraints. We used these predetermined input parameter ranges for each program variable for generating pairwise test suites using {\ctt} in order to maintain the same input model as the one used to create manual test suites.  

\subsection{Evaluation Measurements}
\label{subsec:measurements}
In this section, we present how the case study is conducted with respect to each research question. We first discuss the evaluation measurements used for efficiency and effectiveness of testing.

\subsubsection*{Code Coverage} We use code coverage criteria to assess the test suites thoroughness \cite{ammann2008introduction} and answer RQ1. These coverage criteria are used to evaluate the extent to which the program has been exercised by a certain test suite. In this study, code coverage is directly measured using the branch coverage criterion. For the programs selected in this study the EN50128 safety standard \cite{en200150128} involves achieving high branch coverage. A test suite achieves 100\% branch coverage if executing the program causes each branch in the {\iec} program to have the value \emph{true} and \emph{false} at least once. A branch coverage score was obtained for each generated test suite using our own tool implementation based on the PLC execution framework provided by Bombardier Transportation Sweden AB. 

\subsubsection*{Fault Detection} Ideally, in order to measure fault detection, real faulty versions of the programs are required. In our case, the data provided did not contain any information about what faults occurred during the testing of these programs. To overcome this issue and answer RQ2, we used mutation analysis by generating faulty versions of the original programs. Mutation analysis is a method of automatically creating artificial faulty versions of a program in order to examine the fault detection ability of a test suite \cite{ammann2008introduction}. A \textit{mutant} is a different version of the original program containing a small syntactical change. For example, in an {\iec} program, a mutant is created by replacing a constant value with another one, negating a signal or changing the type of a computational block. If the execution of the resulting mutant on a test is producing a different output as the execution of the original program, the test suite \textit{detects} the mutant. The mutation score is computed using an output-only verdict (i.e., using the expected values for all of the program outputs) against the set of mutants. The fault detection capability of each test suite was calculated as the ratio of mutants detected to the total number of mutants. Just et al. \cite{just2014mutants} provided compelling experimental evidence that the mutation score is a proxy for real fault detection. 

In the creation of mutants we used common type of faults in {\iec} software \cite{oh2005software} as a basis for establishing the following mutation operators:
\begin{itemize}
\item {\it Logic Block Replacement}. Replacing a logical block with another block from the same category (e.g., an OR block is replaced by an AND block).
\item {\it Comparison Block Replacement}. Replacing a comparison block with another block from the same category (e.g., a Greater-Or-Equal (GE) block is replaced by a Greater-Than (GT) block).
\item {\it Arithmetic Block Replacement}. Replacing an arithmetic block with another block from the same category (e.g., replacing a maximum calculation block (MAX) with a minimum calculation block (MIN)).
\item {\it Negation Insertion}. Negating an input or output connection (e.g., an output boolean connection is negated).
\item {\it Value Replacement}. Replacing a value of a constant variable connected to a block (e.g., a constant variable is replaced by its boundary values).
\item {\it Timer Block Replacement}. Replacing a timer block with another block from the same function category (e.g., a Timer-On (TON) block is replaced by a Timer-Off (TOF) block).
\end{itemize}
Each of the mutation operators were applied to each program element. In total, for all of the selected programs, 1597 mutants were generated (i.e., 35 mutants on average per program). A mutant was considered detected by a test suite if the output from the mutated program differed from that of the original program. A mutation score was obtained for each generated test using our own tool implementation. 
%
%
%
%

\begin{figure*}[tbp]
\centering\resizebox{\textwidth}{!}{
\includegraphics[width=0.3\textwidth]{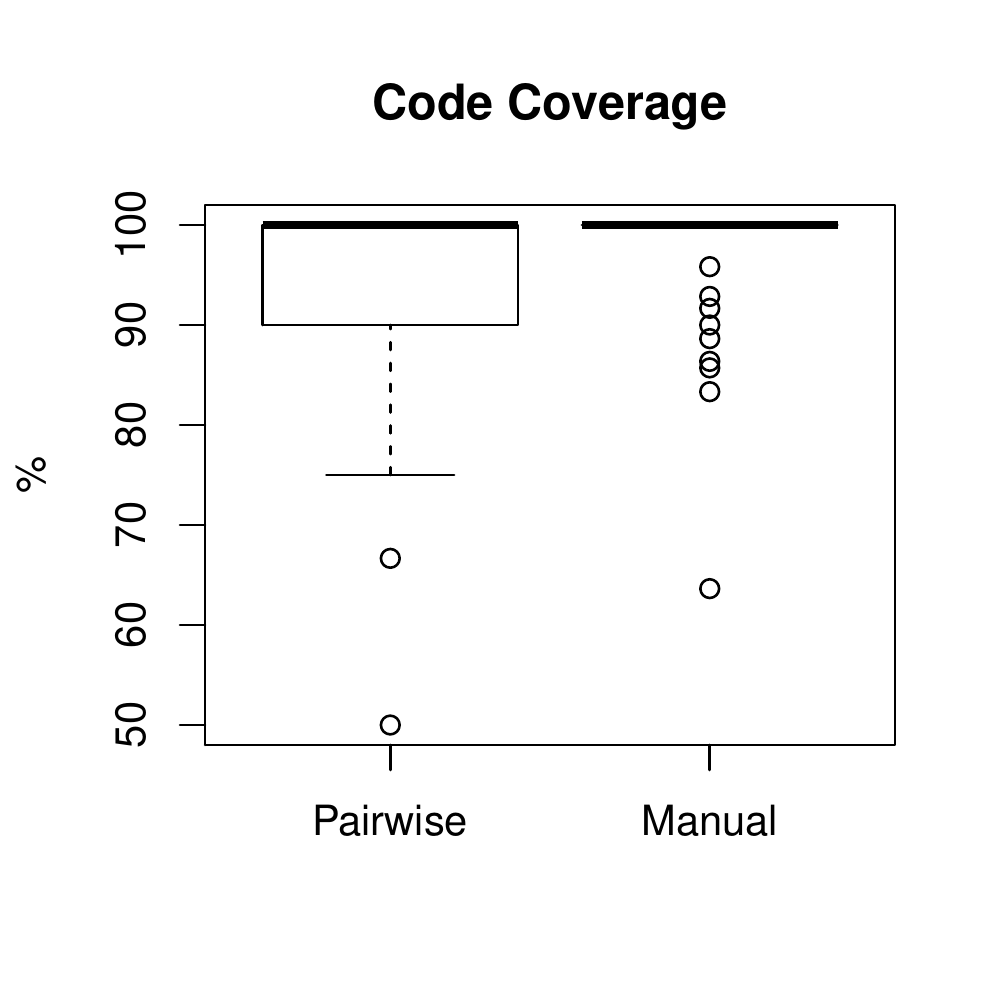}
\includegraphics[width=0.3\textwidth]{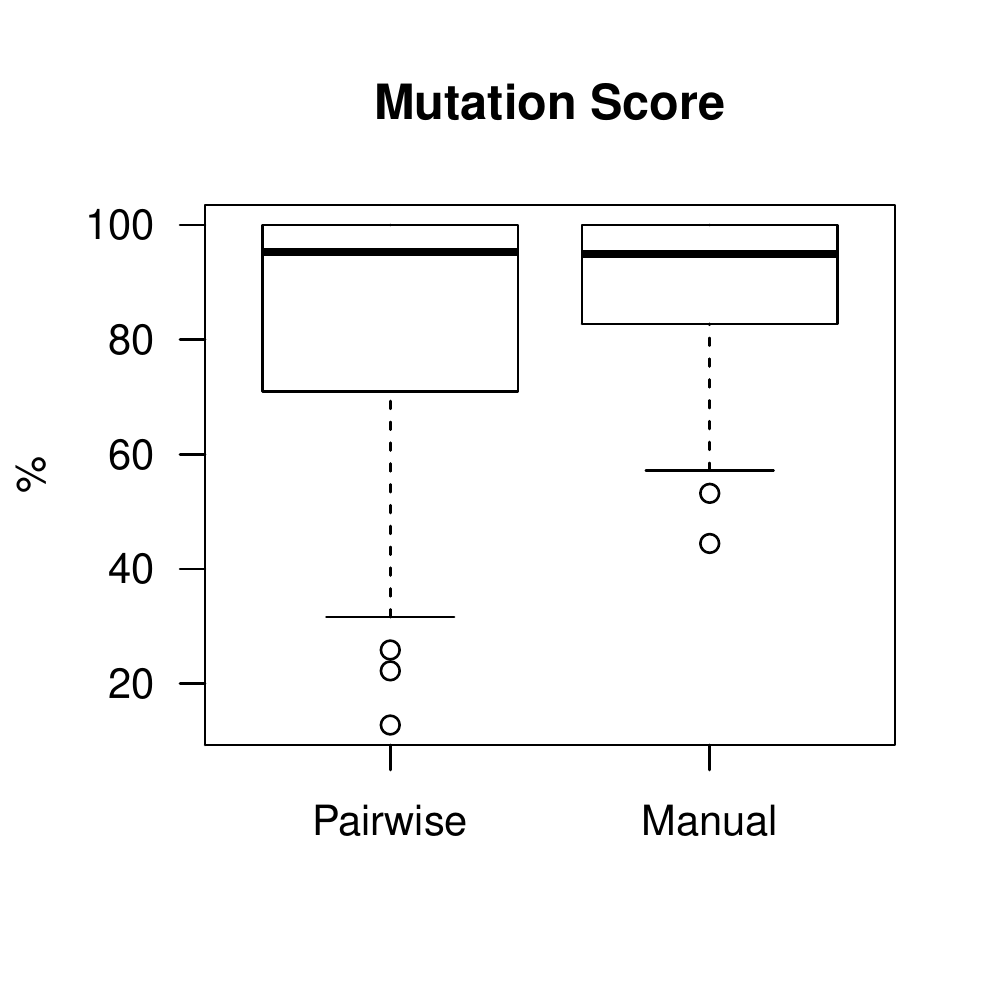}
\includegraphics[width=0.3\textwidth]{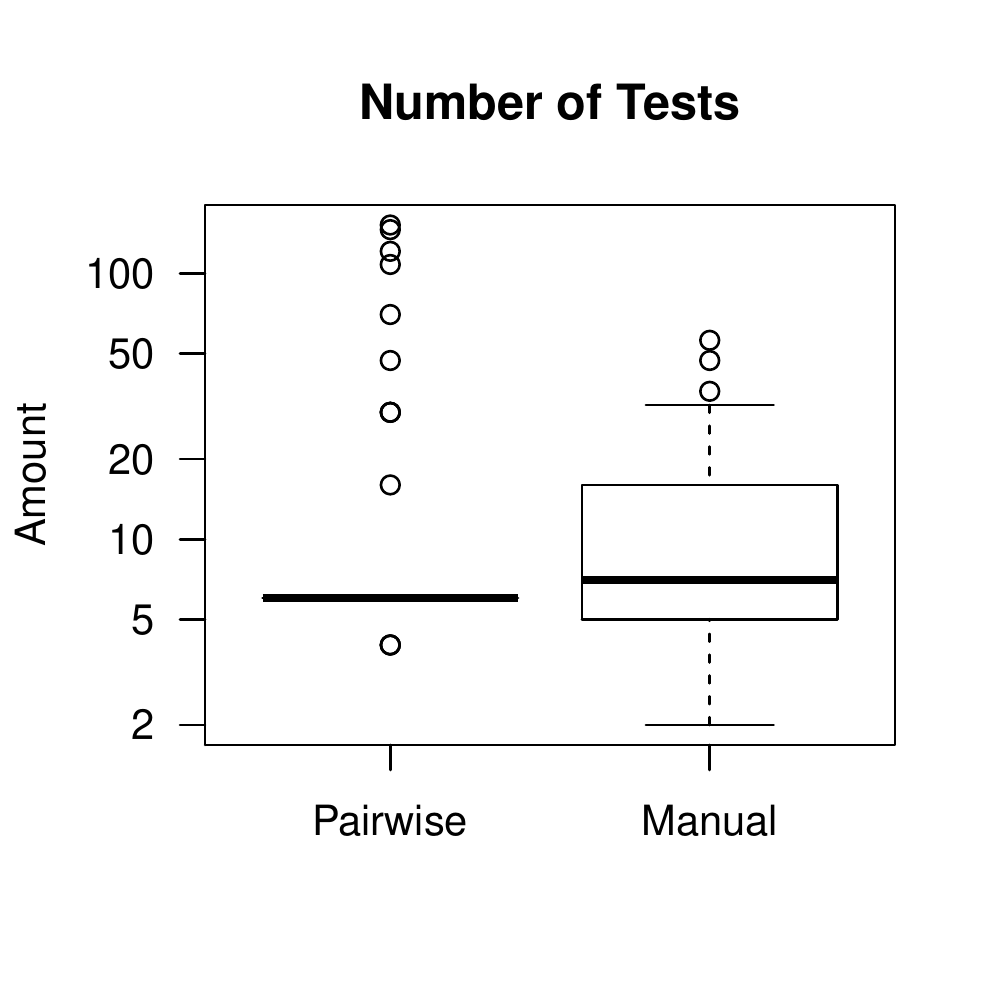}
}
\vspace*{-1cm}
\caption[Box plots]{Code Coverage, mutation score and number of test cases comparison between manually handcrafted test suites (Manual) and test suites generated using pairwise testing (Pairwise); boxes span from 1st to 3rd quartile with black middle lines marking the median and the whiskers extending up to 1.5x the inter-quartile range; the circle symbols represent the outliers.}
\label{figure:boxplots}
\end{figure*}
\begin{figure*}[tbp]
\centering
\resizebox{\textwidth}{!}{
\includegraphics[height=4cm]{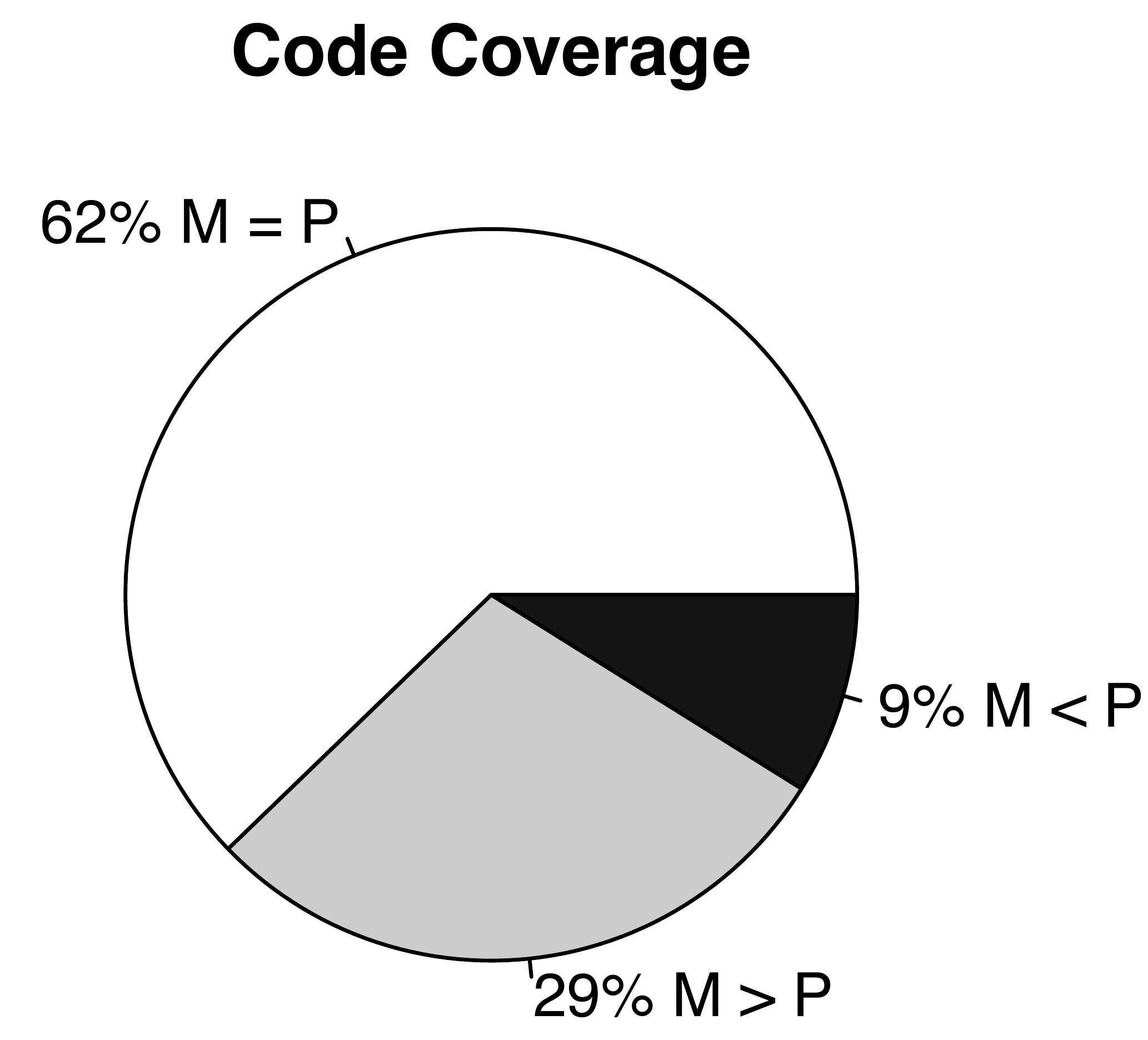}
\hspace{0.7cm}
\includegraphics[height=4cm]{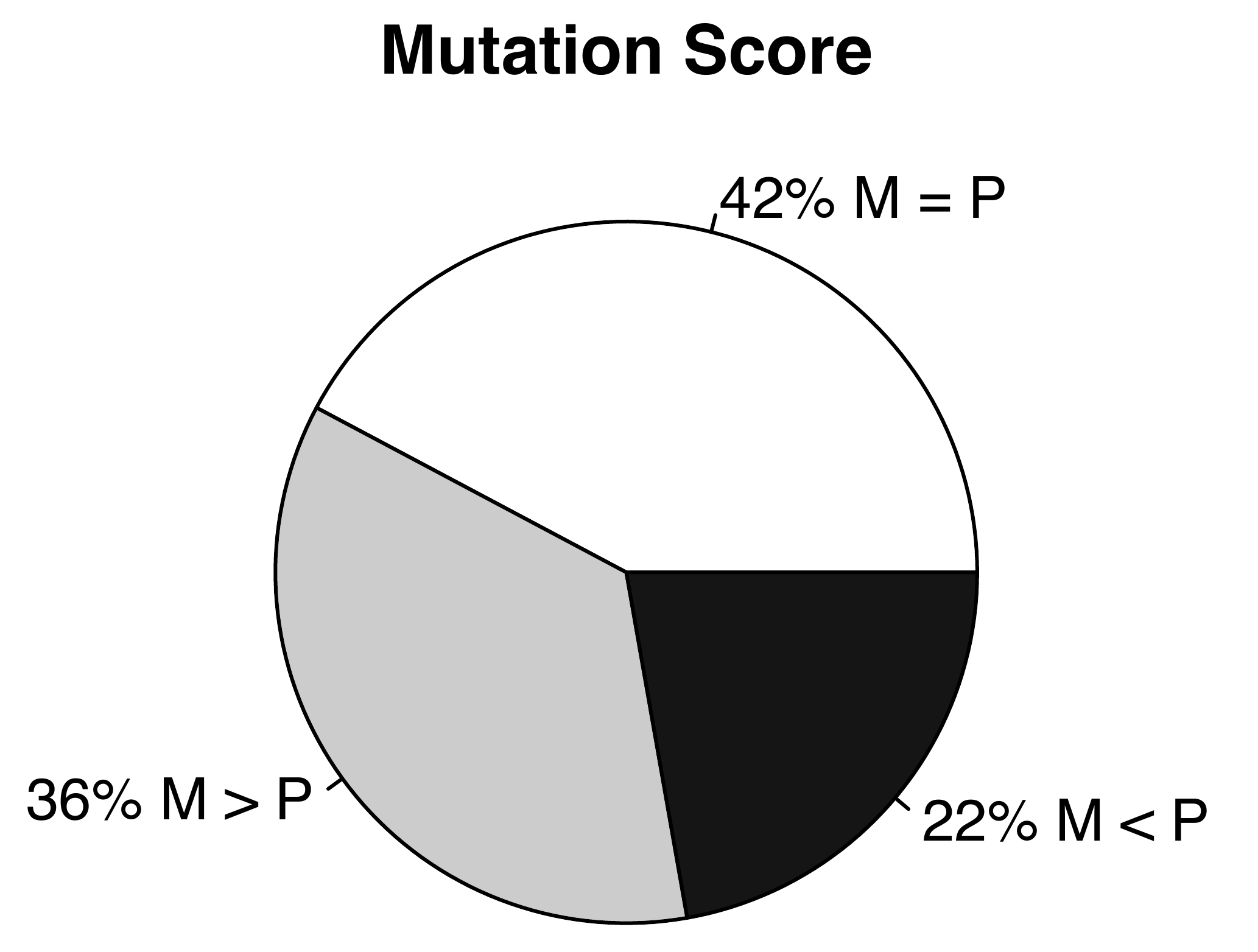}
\hspace{0.8cm}
\includegraphics[height=4cm]{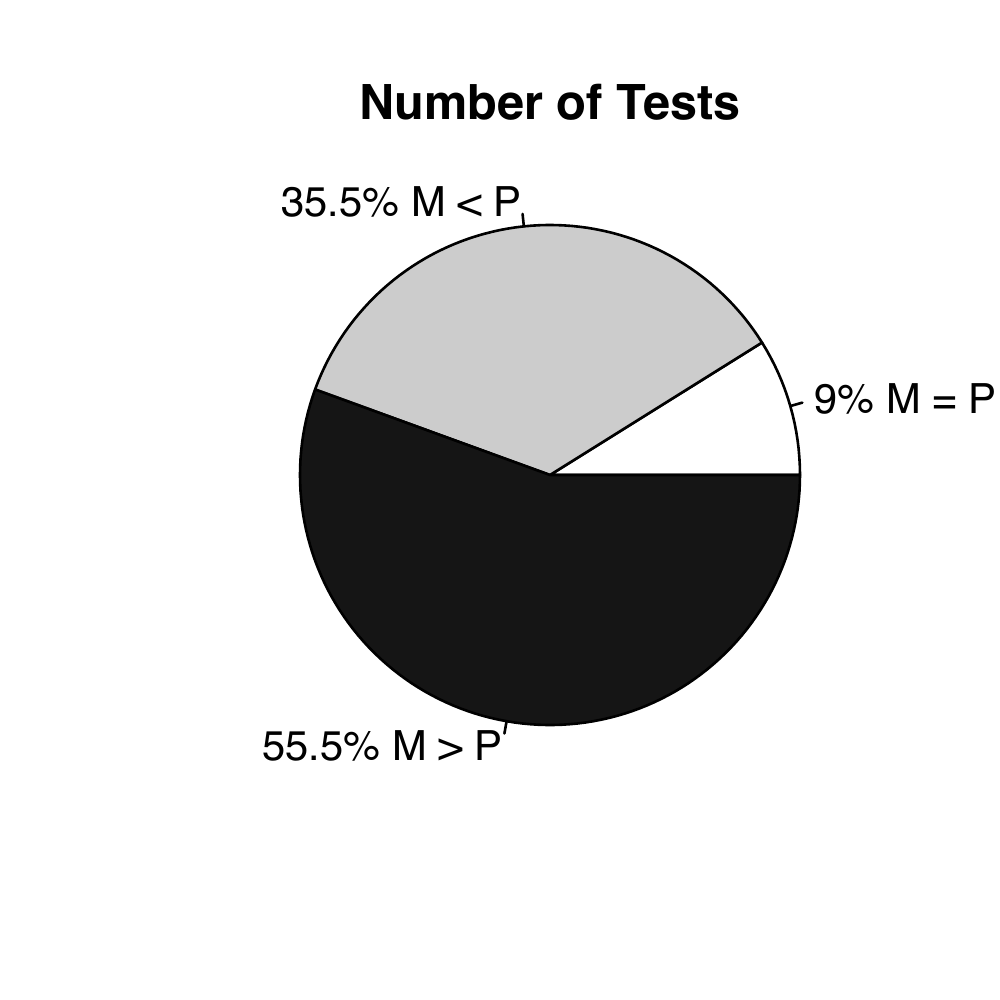}
}
\caption[Manual and Pairwise test suite comparison]{Charts of number of tests, code coverage and mutation scores depicting how each technique (M stands for manual testing and P is short for pairwise testing) compared to the other for each measure. M$>$P indicates that manual testing achieved a higher measurement score, M$<$P shows how many times manual testing achieved a lower measurement score and M$=$P shows when both techniques achieve the same measurement score.}
\label{figure:caseEval}
\end{figure*}

\subsubsection*{Number of Tests} In an ideal situation, the cost of testing is measured by taking into account direct and indirect type of cost, by measuring directly the test suite creation, the test suite execution and the checking of test suite results. However, since this is a case study using programs for which the development was performed a few years back, this kind of cost data was not available. To answer RQ3, we used the number of created test cases as a proxy for efficiency as we assume that all human costs are depended on the number of tests. The higher the number of tests, the higher are the respective testing costs. For example, a complex program will require more effort for test creation, execution and checking of the results.

\section{Experimental Results}
\label{sec:results}
In this section, we quantitatively answer the three research questions posed in Section \ref{sec:method}. We collected the data to answer these research questions by generating test suites for pairwise testing using {\ctt}; collecting manual test suites created by experienced industrial engineers; measuring their code coverage; and measuring their effectiveness in terms of mutation score. The overall results of this study are summarized in the form of boxplots in Figure \ref{figure:boxplots}. In Tables \ref{table:codeCoverage}, \ref{table:mutationScore} and \ref{table:testsuiteSize} we present the code coverage scores, mutation scores and the number of tests in each test suite by listing the standard deviation, mean, median, minimum and maximum values. In addition, statistical analysis was performed using the R statistical software \footnote{https://www.r-project.org/}. We assume that the collected data is drawn from an unknown distribution. In order to evaluate if there is any statistical difference between manual and pairwise testing we use a Wilcoxon-Man-Whitney U-test\cite{howell2012statistical}, a non-parametric hypothesis test used for checking if two data samples are randomly obtained from identical populations. We also use the Vargha-Delaney test (also known as the standardized effect size) to calculate the statistical significance. The Vargha-Delaney $\hat{A}$-measure is also "\textit{a measure of stochastic superiority}"\cite{vargha2000critique} and is used to measure the difference between two populations. The test result is denoted as $\hat{A}$, and simply specifies the amount of times population A is expected to be better than population B\cite{neumann2015transformed}. Its significance is determined when the effect size is above 0.7 or below 0.2.

\subsection{Code Coverage}
\label{subsec:code}
RQ1 asked if pairwise testing achieves better code coverage scores than manual testing. The coverage scores achieved by pairwise testing are ranging between 50\% and 100\% while for manual testing these are varying between 63\% and 100\%. As shown in Table \ref{table:codeCoverage}, the use of manual testing achieves on average 97\% branch coverage (3\% on average higher than pairwise testing). Results for all programs (in Table \ref{table:statisticalAnalysis}) show that differences in code coverage between manual and pairwise testing are statistically significant with a p-value of 0.04 but their effect is not strong (i.e., an effect size of 0.6).

\begin{table}[tbp]
\centering
\renewcommand{\arraystretch}{1.5}
\caption[Overview of each test methods results in regards to Code Coverage]{Results for code coverage measurements.
}
\resizebox{\columnwidth}{!}{
\begin{tabular}{l|r|r|r|r|r|}
\cline{2-6}
Code Coverage                  & \multicolumn{1}{c|}{SD} & \multicolumn{1}{c|}{Mean} & \multicolumn{1}{c|}{Median} & \multicolumn{1}{c|}{MIN} & \multicolumn{1}{c|}{MAX} \\ \hline
\multicolumn{1}{|l|}{Pairwise} & 10.35                   & 93.95                     & 100.00                      & 50.00                    & 100.00                   \\ \hline
\multicolumn{1}{|l|}{Manual} & 6.71                    & 97.29                     & 100.00                      & 63.64                    & 100.00                   \\ \hline
\end{tabular}
}
\label{table:codeCoverage}
\end{table}

As seen in Figure \ref{figure:caseEval}, for 62\% of the programs considered, pairwise performs equally good as manual testing; for 29\% of the programs manual testing performed better in terms of achieved code coverage while for 9\% of the programs pairwise testing covers more code than manual testing.

The results for all programs were surprising: test suites created using pairwise testing achieved relatively high code coverage (94\% on average). This shows that, for the programs studied in this experiment, pairwise testing achieves high branch coverage. This is likely due to the complexity of the studied programs. It is possible that more complex software would yield a greater code coverage difference between manual and pairwise test suites.  

Overall, as shown in Figure \ref{figure:caseEval}, we confirm that pairwise test suites achieve just as good or better code coverage scores as manual testing for 71\% of programs considered in this study. This can be explained by the fact that pairwise testing if properly used is quite good at covering the logical behaviour of the code.

\begin{mdframed}[style=style1]
{\it Answer RQ1: Code coverage scores achieved by pairwise test suites are slightly lower than the ones created manually by industrial engineers.}
\end{mdframed}

\begin{table}[tbp]
\centering
\renewcommand{\arraystretch}{1.5}
\caption[Overview of each test methods results in regards to mutation score]{Results for fault detection (mutation analysis) measurements.
}
\resizebox{\columnwidth}{!}{
\begin{tabular}{l|r|r|r|r|r|}
\cline{2-6}
Mutation Score                & \multicolumn{1}{c|}{SD} & \multicolumn{1}{c|}{Mean} & \multicolumn{1}{c|}{Median} & \multicolumn{1}{c|}{MIN} & \multicolumn{1}{c|}{MAX} \\ \hline
\multicolumn{1}{|l|}{Pairwise} & 25.69                   & 81.58                     & 95.24                       & 12.77                    & 100.00                   \\ \hline 
\multicolumn{1}{|l|}{Manual} & 14.22                   & 88.90                     & 95.00                       & 44.44                    & 100.00                   \\ \hline 
\end{tabular}
}
\label{table:mutationScore}
\end{table}
\subsection{Fault Detection}
\label{subsec:fault}
To answer RQ2, we first computed the mutation score of each manual and pairwise test suites. Figure \ref{figure:boxplots} shows box plots of our results for fault detection in terms of mutation score. Table \ref{table:mutationScore} summarizes statistics for these test suites. For all programs the fault detection scores obtained by manually written test suites are higher on average with 7\% than those achieved by pairwise testing. However, there is no statistically significant difference at 0.05; as the p-value is 0.67 and the effect size is 0.53 in Table \ref{table:statisticalAnalysis}. A larger sample size would be needed to obtain more confidence in our results. Interestingly, as show in Figure \ref{figure:caseEval}, our results suggest that fault detection scores achieved by manual testing are not significantly better at finding faults than pairwise testing. It seems that test suites generated using pairwise testing are just as good in terms of fault detection as manual test suites for 64\% of the cases considered in this study. For 42\% of the programs, pairwise testing performs as well as manual testing while for 36\% of the programs manual testing performed better in terms of fault detection. 

The difference in effectiveness between manual and pairwise could be due to other factors such as the number of test cases and the test design techniques used to manually create test suites (e.g., testing the timed behavior of the PLC software). 

\begin{mdframed}[style=style1]
{\it Answer RQ2: Pairwise testing is able to produce comparable results to manual testing in terms of fault detection. However, manual testing produced better mutation scores on average.}
\end{mdframed}


\subsection{Number of Tests}
\label{subsec:size}
\begin{table}[tbp]
\centering
\renewcommand{\arraystretch}{1.5}
\caption[Overview of each test methods results in regards to test suite size]{Results for code number of tests (test suite size) measurements.
}
\resizebox{\columnwidth}{!}{
\begin{tabular}{l|r|r|r|r|r|}
\cline{2-6}
Number of Tests  							& \multicolumn{1}{c|}{SD} 	& \multicolumn{1}{c|}{Mean} 	& \multicolumn{1}{c|}{Median} 	& \multicolumn{1}{c|}{MIN} & \multicolumn{1}{c|}{MAX} \\ \hline
\multicolumn{1}{|l|}{Pairwise} 				& 37.05                     & 21.20                   		& 6.00                     		& 4.00                     & 152.00                   \\ \hline
\multicolumn{1}{|l|}{Manual}				& 12.26                     & 12.98                   		& 7.00                     		& 2.00                     & 56.00                    \\ \hline
\end{tabular}
}
\label{table:testsuiteSize}
\end{table}

This section aims to answer RQ3 regarding the relative cost of performing manual testing versus pairwise testing. Analysing the cost in this study is directly related to the number of test cases giving a picture of the effort needed per created test suite. Based on the results highlighted in Figure \ref{figure:boxplots}, the use of pairwise testing results in very inconsistent number of tests created, compared to manual testing which seems to create tests with more diverse number of steps than pairwise testing. Examining Table \ref{table:testsuiteSize}, we see a different pattern: less number of tests on average are manually created by industrial engineers (13 test cases on average in a test suite) than when using pairwise testing (21 test cases on average in a test suite). As seen in Figure \ref{figure:caseEval}, for 44\% of the programs considered, pairwise test suites produced equally or larger test suites than the manual ones, leading to 55\% where pairwise produced fewer. Table \ref{table:statisticalAnalysis} shows an interesting pattern in the statistical analysis: the standardized effect size being 0.53, with p-value being higher than the traditional statistical significance limit of 0.05. Results are not strong in terms of effect size and we did not obtain any statistical difference for the number of tests. 

The results for all programs matched our expectations: manual tests are handcrafted by experienced industrial engineers that can create very diverse tests. It is possible that more complex software would yield greater number of tests differences between tests written manually and pairwise testing. 
\begin{mdframed}[style=style1]
{\it Answer RQ3: The use of pairwise testing results in more number of tests created on average than the use of manual testing. Even so, pairwise testing produced more cases with less tests created compared to manual testing, resulting in very inconsistent results.}
\end{mdframed}



\begin{table}[tpb]
\centering
\renewcommand{\arraystretch}{1.5}
\caption[Statistical Analysis of the results]{Results of the statistical analysis. For each metric we calculated the effect size of each test creation method compared to each other. We also report the p-values of a Wilcoxon-Mann-Whitney U-tests.} 
\resizebox{\columnwidth}{!}{ 
\begin{tabular}{l|l|r|r|r|}
\cline{2-5}
Statistics & Values & \multicolumn{1}{l|}{Code Coverage} & \multicolumn{1}{l|}{Mutation Score} & \multicolumn{1}{l|}{Test Suite Size} \\ \hline
\multicolumn{1}{|l|}{\multirow{2}{*}{M vs P}}    & $\vcenter{\hbox{$\hat{A}$}}$ & 0.60                                 & 0.53                               & 0.53                               \\ \cline{2-5} 
\multicolumn{1}{|l|}{}                  				& $p-value$ 	& 0.04                                 & 0.67                           & 0.64                            \\ \hline
\end{tabular}
}
\label{table:statisticalAnalysis}
\end{table}

\section{Discussion}
\label{sec:discussion}
The goal of this work was to compare pairwise testing with manual testing performed by industrial engineers in terms of code coverage, fault detection and the number of created test cases. We found out that pairwise testing achieves high code coverage, but slightly lower scores than manual testing. In addition, we found the fault detection scores for pairwise testing to be lower on average than the ones written manually by industrial engineers. Interestingly, pairwise testing achieves equally good or better fault detection scores than manual testing for 64\% of the programs considered, which might indicate that for more than half of the programs the fault detection scores for pairwise testing are a good predictor of test effectiveness. This is reinforced by the mutation score box plot in Figure \ref{figure:boxplots}: the median mutation score for pairwise testing is 95\%. This in combination with the achieved high code coverage, suggests that pairwise testing can cover as much code as manual testing performed by industrial engineers, but this is not entirely reflected in its fault detection capability. The fault detection rate between manual and pairwise testing was found, in some of the published studies \cite{ellims2008effectiveness,ellims2007aetg} to be similar to our results. Interestingly enough, our results indicate that pairwise test suites might be even better in some cases in terms of fault detection than manual test suites. However, a larger pool of programs and tests is needed to statistically confirm this hypothesis.

The mean value for fault detection of 81\% for pairwise testing is right in line with the proportion of 2-way faults seen in other domains \cite{ kuhn20132combinatorial}. It is interesting to note here that the fault distribution for PLC industrial control software is similar to other types of software.

As part of our study, we used the number of tests to estimate the test efficiency in terms of creation, execution and result checking. While the cost of creating and executing a test for pairwise testing can be low compared to manual testing, the cost of checking the result is usually human intensive. Practically, the higher the number of test cases, the higher the cost of checking the test result. Our study suggests that pairwise test suites, while inconsistent, are longer on average in terms of created test steps (number of tests) than manual test suites. By considering generating optimized or shorter test suites, one could improve the cost of performing pairwise testing. 

The idea of using pairwise testing in practice stands as a significant progress in the development of automatic test generation approaches. This progress implies, to some extent, that pairwise testing should be at least as effective and more efficient than manual testing for it to be considered ready to be used as a replacement to the manual effort of creating tests. The overall result, from this case study, is that pairwise testing alone is not better than manual testing. However, pairwise testing or stronger combinatorial criteria are capable of at least aiding an engineer in testing of industrial software. Our observations showed that experienced engineers are very effective at generating the right choice of values and considering the timing of the input parameters. Industrial PLC software typically have a complex and time-dependent behaviour. This behaviour require inputs to retain and change a sequence of inputs for some time in order to trigger a certain logical event. Several manual test suites collected in this case study contained that kind of test cases. The engineers creating these test suites had years of experience in developing and testing this type of software, including good knowledge of what combinatorial interactions are needed to cover the code and detect faults. As it turns out, pairwise testing is not particularly useful for some of the programs considered in this study. By considering generating more complex (stronger t-wise) and time-depended tests, one could improve both the achieved code coverage as well as the fault detection capability.

\section{Threats to Validity}
\label{sec:validity}

There are many tools (e.g., ACTS \cite{yu2013acts}, AETG\cite{cohen1997aetg}, TCG\cite{tung2000automating}) for automatically generating tests using pairwise testing and these may produce different results. The use of these tools in this study is complicated by the modelling of the input space for a PLC program. Hence, we choose a tool specifically developed for testing PLC industrial programs. For more details on the comparison between {\ctt} and ACTS, we refer the reader to the work of Charbachi and Eklund \cite{thesischa}. {\ctt} is using the IPOG algorithm. This algorithm continually expands the test suite to fit the input parameters with the use of vertical and horizontal extension. {\ctt} currently uses a first element tie breaker which was chosen when a clean-cut best choice for tie breaking was absent \cite{huang2014tie}. This choice of a tie breaker might affect the effectiveness of the generated tests. 

Another threat to the validity of this study is also related to the use of the {\ctt} tool. Vertical extension \cite{lei2007ipog} is a step which adds new tests, if needed, to the test set produced by horizontal growth when there are no modifiable test cases to cover a specific pair. The IPOG algorithm, in this case, creates a new test case, thus expanding the size of the test suite by making the test case to cover the specific pair and keeping all the other parameters modifiable. This is done to reduce the need of further vertical extensions and can be used to avoid creating a new test case. However, when trying to execute a test suite, these values are syntactically illegal and need to be changed. {\ctt} currently handles this by randomising the modifiable values to a default option for each parameter. A more accurate option for each parameter would be needed to obtain more confidence in the test suite effectiveness.

The data collected is based on a study in one company using one industrial system containing 45 programs and manual test suites created by industrial engineers. We argue that even if the number of programs can be considered relatively small, reporting a case study using industrial artefacts can be representative.  

Since this case study was performed post-mortem, the cost information was not available. We used the number of test cases as a valid proxy measure and a more detailed cost model should be used to obtain more accurate results.


\section{Conclusion}
\label{sec:conclusions}
In this paper, we studied the comparison between pairwise testing and manual testing in terms of branch coverage, mutation score and number of created test cases. From the 45 PLC industrial programs we studied, we drew the following conclusions:
\begin{itemize}
\item The use of pairwise testing results in high branch coverage and mutation scores for the majority of the programs considered. 
\item The results of this paper support the claim that pairwise testing is not quite as effective (i.e., achieved branch coverage and fault detection) and efficient (in terms of number of tests created) as manual testing. 
\item The use of pairwise testing results in similar or better fault detection scores than the use of manual testing for 64\% of the programs.
\end{itemize}

The results imply that pairwise testing can achieve high branch coverage, but slightly lower scores than manual testing.
In summary, our results suggests that pairwise testing can perform in some cases comparably with manual testing performed by industrial engineers. This is a significant experimental evidence on the progress of pairwise testing that needs to be further studied; we need to consider the cost of using pairwise testing in practice. In addition, pairwise testing is only one type of combination strategy and we would need to evaluate the use of stronger criteria such as t-wise (with $t>2$) testing. 


%

\ifCLASSOPTIONcompsoc
  \section*{Acknowledgments}
\else
  \section*{Acknowledgment}
\fi
This research was supported by the Swedish Research Council (VR) through the “Adequacy-based testing of extra-functional properties of embedded systems” project and by VINNOVA and ECSEL (EU's Horizon 2020) under grant agreement No 737494. The authors would like to thank Bombardier Transportation AB for the valuable assistance in the planning and execution of this work. 

\ifCLASSOPTIONcaptionsoff
  \newpage
\fi




\balance
%
\bibliographystyle{IEEEtran}
\bibliography{./references}



%






\end{document}